\documentclass{iopart}
\usepackage[english]{babel}
\usepackage[dvips]{graphicx}
\usepackage{amssymb}
\usepackage{amsthm}
\usepackage{cite}
\usepackage{color}
\usepackage{iopams}
\usepackage{latexsym}
\usepackage{times}

\newcommand{\cf}{\textit{cf.}~}
\newcommand{\ie}{\textit{i.e.}~}
\newcommand{\eg}{\textit{e.g.}~}

\begin{document}
\title[Critical Phenomena in Neutron Stars I: Linearly Unstable
  Nonrotating Models] {Critical Phenomena in Neutron Stars I: Linearly
  Unstable Nonrotating Models}

\author{David Radice$^1$, Luciano Rezzolla$^{1,2}$ and Thorsten Kellerman$^1$}

\address{$^1$ Max-Planck-Institut f\"ur Gravitationsphysik, Albert Einstein
Institut, Potsdam, Germany}

\address{$^2$ Department of Physics and Astronomy, Louisiana State
  University, Baton Rouge, USA}

\begin{abstract}
We consider the evolution in full general relativity of a family of
linearly \emph{unstable} isolated spherical neutron stars under the
effects of very small, perturbations as induced by the truncation
error. Using a simple ideal-fluid equation of state we find that this
system exhibits a type-I critical behaviour, thus confirming the
conclusions reached by Liebling et al.~\cite{liebling_2010_emr} for
rotating magnetized stars. Exploiting the relative simplicity of our
system, we are able carry out a more in-depth study providing solid
evidences of the criticality of this phenomenon and also to give a
simple interpretation of the putative critical solution as a spherical
solution with the unstable mode being the fundamental F-mode. Hence
for any choice of the polytropic constant, the critical solution will
distinguish the set of subcritical models migrating to the stable
branch of the models of equilibrium from the set of subcritical models
collapsing to a black hole. Finally, we study how the dynamics changes
when the numerically perturbation is replaced by a finite-size,
resolution independent velocity perturbation and show that in such
cases a nearly-critical solution can be changed into either a sub or
supercritical. The work reported here also lays the basis for the
analysis carried in a companion paper, where the critical behaviour in
the head-on collision of two neutron stars is instead
considered~\cite{Kellermann:10}.
\end{abstract}

\pacs{
04.25.Dm, 
04.40.Dg, 
04.70.Bw, 
95.30.Lz, 
97.60.Jd
}

\section{Introduction}

Critical phenomena in general relativity were first discovered by
Choptuik~\cite{Choptuik93} for the gravitational collapse of a
massless scalar field. After his seminal work these phenomena were
also discovered for a wide range of systems, including massive scalar
fields and ultra-relativistic fluids~\cite{evans_1994_cps, Hara96a,
  neilsen_2000_cpp, novak_2001_vic, noble_2003_nsr, Noble08a, Jin:07a,
  musco05, musco09} (see also~\cite{gundlach_2007_cpg} for a recent
review). Our interest here is to further develop the analysis of
critical phenomena in the presence of (perfect) fluids in general and
of relativistic spherical stars in particular.

We recall that the first simulations of critical collapse of a perfect
fluid were performed by Evans and Coleman~\cite{evans_1994_cps}. They
used an ultra-relativistic equation of state (EOS) with polytropic
exponent $\Gamma=4/3$ (radiation) and found evidence of a continuously
self-similar (CSS) critical solution. Following works by
\cite{Hara96a} and~\cite{brady_2002_bht} showed that type-II critical
phenomena occur for every value of $\Gamma$ between $1$ and $2$ with a
CSS critical solution.  The ultra-relativistic EOS, that is one in
which the pressure is proportional to the energy density $p=(\Gamma -
1)e$, is the only scale-invariant equation of state in general
relativity, thus the only one compatible with the existence of an
homothetic vector field. In other words, fluids obeying an
ultra-relativistic EOS are the ones that admit self-similar solutions
in general relativity~\cite{Cahill70}. Nevertheless evidence of CSS
critical solutions were found by~\cite{neilsen_2000_cpp} also for
perfect fluids with an ideal-gas EOS. They conjectured that, as
type-II phenomena are kinetically dominated, an ideal gas would behave
like an ultra-relativistic gas and thus nearly-critical solutions will
approach the corresponding ultra-relativistic CSS solution. They were
actually able to show that this is indeed the case by comparing
ultra-relativistic CSS solutions, obtained using the self-similarity
ansatz, to the corresponding ideal-gas solution. This suggested the
possibility of observing type II critical phenomena in the context of
neutron star collapse.

The first studies in this direction were performed by Novak
\cite{novak_2001_vic}. He studied solutions with velocity induced
perturbation of a Tolman-Oppenheimer-Volkoff (TOV) background and
found a type-II critical phenomena using a stiff ideal gas EOS (\ie
with $\Gamma = 2$) and even a more realistic tabulated
EOS. Noble~\cite{noble_2003_nsr} studied critical phenomena in neutron
star collapse in great detail by using the gravitational interaction
with a massless scalar field to perturb spherical-star
configurations. He confirmed the observation by Novak that a minimum
mass is required to trigger collapse and also found that, for very
high mass spherical stars, type-I critical phenomena is observed with
each model oscillating around a corresponding solution on the unstable
branch.  More recently~\cite{Noble08a} performed an accurate study
using the same setup of~\cite{novak_2001_vic} and was able to show
that, even for initial data of spherical stars, the scaling exponent
in the mass relation law is compatible with the exponent in the
corresponding ultra-relativistic case.

Type-I critical phenomena in neutron star collapse have seen a renewed
interested mainly due to the discover by~\cite{Jin:07a} of critical
phenomena in head-on collision of neutron stars (NS). They considered
families of two equal mass NS, modeled with an ideal gas EOS, boosted
towards each other at a prescribed speed and varied the mass of the
stars, their separation, velocity and $\Gamma$-parameter. They found
that at the threshold of black hole formation a type-I critical
phenomena can be observed, with the putative solution being an
oscillating star. In a successive paper~\cite{wan_2008_das} repeated
the same results by performing head-on collision of Gaussian
packets. They also claimed that the critical solution is a new kind of
metastable object and not a perturbed spherical star as its mass is
less then the maximum allowed mass of a spherical star for the same
EOS~\cite{Jin:07a} and its oscillation frequencies are one order of
magnitude larger then the eigenfrequencies of a spherical star with
the same total baryonic mass \cite{wan_2008_das}. This claims have
partly been rejected by the work carried out in~\cite{Kellermann:10},
which also investigated critical phenomena in head-on collision of
equal-mass spherical stars and found that the critical solution can be
compared to a spherical star solution sitting on the unstable branch.

Motivated by these recent results we have investigated phase
transitions in families of linearly unstable spherical stars, a case
substantially unexplored in previous systematic works. We show that
this transition is a critical one and that the putative critical
solution can be interpreted as an oscillating spherical star. We
further study the effects of the introduction of velocity-induced
perturbations, as the ones used by \cite{novak_2001_vic}, on nearly
critical solutions.

The remainder of this paper is organized as follows. In section
\ref{sec:critical.phenomena} we give a brief introduction to critical
phenomena in general relativity. In section \ref{sec:numerical.setup}
we describe the numerical settings of the simulations and the
properties of the used initial data. In section \ref{sec:results} we
show in details our results, while section \ref{sec:conclusions} is
dedicated to conclusions and discussion. We use a spacetime signature
$(-,+,+,+)$, with Greek indices running from 0 to 3 and Latin indices
from 1 to 3. We also employ the standard convention for the summation
over repeated indices. Unless otherwise stated, all the quantities are
expressed in a system of units in which $c=G=M_\odot=1$.
\section{Critical phenomena in gravitational collapse}
\label{sec:critical.phenomena}

In what follows we give a brief overview of critical phenomena in
gravitational collapse and which will be useful to cast our results in
the more general context of critical phenomena in general
relativity. We refer the interested reader to~\cite{gundlach_2007_cpg}
for a more systematic presentation.

\subsection{Self-similarity}

Before dwelling on critical phenomena and because self-similarity
plays a central role in this context, it is useful to recall briefly
the definitions of ``continuous'' self-similarity and ``discrete''
self-similarity. We refer the interested reader
to~\cite{gundlach_1999_cpg} for a more detailed discussion.

We recall that a spacetime is said to be continuously self-similar if
there exist a vector field, $\xi^\mu$ such that $\nabla_{(\mu}
\xi_{\nu)} = g_{\mu\nu}$. Vector fields satisfying this condition are
said to be ``homothetic'' as we can easily construct a one-parameter
group of transformations, $\phi_s\colon x^\mu \mapsto y^\mu(s)$, where
$y^\mu(s)$ is the integral curve associated with $\xi^\mu$ passing
through $x^\mu$. It is then easy to see that $\phi_s$ is an homothetic
transformation as the associated push-forward, acts as a rescaling on
the metric
\begin{equation}\label{eq:CSS}
	\phi_s^* g_{\mu\nu} = e^{2s} g_{\mu\nu}\,.
\end{equation}
For this reason in a system of coordinates adapted to the
self-similarity
\begin{equation}
	\xi^\mu = - \bigg( \frac{\partial}{\partial\tau} \bigg)^\mu\,,
\end{equation}
the metric coefficients read
\begin{equation}
	g_{\mu\nu}(\tau,x^i) = e^{-2\tau} \tilde{g}_{\mu\nu}(x^i)\,,
\end{equation}
and the new metric $\tilde{g}_{\mu\nu}$ appears explicitly
self-similar, \ie independent of the $\tau$.

Similarly, a spacetime is said to be ``discretely self-similar'' (DSS) if a
discrete version of (\ref{eq:CSS}) holds. In particular,
Gundlach~\cite{Gundlach97f} defines a spacetime to be DSS if there exist a
diffeomorphism $\phi$ and a real constant $\Delta$ such that for any positive
integer $n$
\begin{equation}\label{eq:DSS}
	(\phi^*)^n g_{\mu\nu} = e^{2 n \Delta} g_{\mu\nu}\,.
\end{equation}
In coordinates adapted to the self-similarity a point $P$ with
coordinates $(\tau,x^i)$ is mapped by $\phi$ into $(\tau-\Delta,x^i)$
and the metric can be written as
\begin{equation}
	g_{\mu\nu}(\tau,x^i) = e^{-2 \tau} \tilde{g}_{\mu\nu}(\tau,x^i)\,,
\end{equation}
where
\begin{equation}
	\tilde{g}_{\mu\nu}(\tau+\Delta,x^i) = \tilde{g}_{\mu\nu}(\tau,x^i)\,.
\end{equation}
Thus, if $\nabla^\mu \tau$ is timelike and induces a Cauchy foliation
of the spacetime, we can give a physical interpretation of the
dynamics of DSS solutions as a combined effect of a rescaling and a
periodic ``echoing'' of the geometry.

\begin{figure}
\begin{center}
	\includegraphics[scale=0.375,angle=-90]{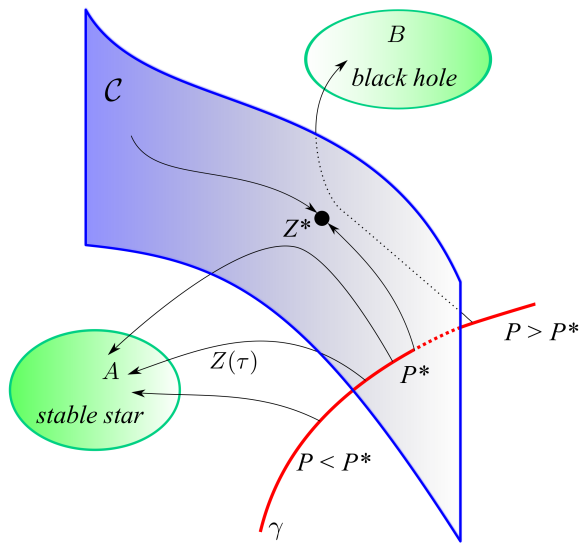}
	\caption{Phase space picture of type-II critical
          phenomena. The surface $\mathcal{C}$ represents the critical
          manifold, separating the basins of attraction of $A$ and
          $B$. The line $\gamma$ represents a generic one-parameter
          family of initial data intersecting the critical manifold in
          $P^\star$. Generic initial data starting at $Z(0)$ will
          evolve towards $A$ or $B$ following the arrows $Z(\tau)$,
          data near the threshold will be marginally attracted towards
          the critical solution $Z^\star$. Points exactly on the
          critical manifold will be attracted to the critical
          solution.\label{fig:phase.space.type.2}}
	\end{center}
\end{figure}

\subsection{The basic concepts}

Let us consider a group of one-parameter families of solutions,
$\mathcal{S}[P]$, of the Einstein equations such that for every $P >
P^\star$, $\mathcal{S}[P]$ contains a black hole and for every $P <
P^\star$, $\mathcal{S}[P]$ is a solution not containing
singularities. We say that these families exhibit a critical
phenomenon if they have the common property that, as $P$ approaches
$P^\star$, $\mathcal{S}[P]$ approaches a universal solution
$\mathcal{S}[P^\star]$, \ie not depending on the particular family of
initial data, and that all the physical quantities of $\mathcal{S}[P]$
depend only on $|P-P^\star|$. In analogy with critical phase
transitions in statistical mechanics, these phenomena are then
classified as type-II or type-I critical
phenomena~\cite{gundlach_2007_cpg}. In what follows we briefly recall
the differences between the two classes.

\subsubsection{Type-II critical phenomena.}
Type-II critical phenomena involve the existence of a CSS or DSS
solutions sitting at the threshold of black-hole formation.  They are
characterized by the mass-scaling relation:
\begin{equation}\label{eq:critphen.type2.fond.relation}
		M_{\mathrm{BH}} = c | P - P^\star |^\gamma\, ,
\end{equation}
where $\gamma$ is independent upon the particular choice of the
initial data. The nomenclature ``type-II'' comes from the analogous
type-II phase transitions in statistical mechanics, which are
characterized by scale invariance of the thermodynamical
quantities~\cite{gundlach_2007_cpg}.

These phenomena are usually interpreted in terms of attractors in an
infinite-dimensional phase space, but we will here present a
qualitative picture which can be useful to fix the ideas (see also the
review in~\cite{gundlach_2007_cpg}). A more rigorous study employing
the renormalization group formalism can be found instead
in~\cite{Hara96a}.

Let us consider general relativity as an infinite dimensional
dynamical system in an abstract phase space in which extra gauge
freedoms have been eliminated so that each point, $Z$, can be thought
as an initial data-set for the Einstein equations and the associated
time development as a line in this space: $t\mapsto Z(t)$. We suppose
to have chosen a slicing adapted to the self-similarity of the
critical solution so that it appears as a fixed point, $Z^\star$, in
the CSS case or a closed orbit for the DSS case
(see~\cite{gundlach_2007_cpg} for a more in-depth discussion of the
consequences of these assumptions).

In the case of CSS solutions, the main features of this phase space
are the presence of two attractive sets: $A$ and $B$ representing
regular solutions without singularities and black-hole
solutions. Their basins of attractions are separated by a manifold,
$\mathcal{C}$, called \emph{critical manifold} on which there is an
attractor of codimension one: the critical solution, $Z^\star$; this
is shown schematically in figure~\ref{fig:phase.space.type.2}.  Any
generic one-parameter family of initial data can then be thought as a
1-dimensional line intersecting the critical manifold in one
point. Initial data with $P<P^\star$, will develop as regular
solutions not containing singularities and will therefore fall in the
basin of attraction containing the so called \emph{subcritical
  solutions} (\cf set $A$ in figure~\ref{fig:phase.space.type.2}).
Conversely, solutions with $P>P^\star$ will undergo gravitational
collapse with the formation of a black hole, thus falling in the basin
of attraction containing the so called \emph{supercritical solutions}
(\cf set $B$ in figure~\ref{fig:phase.space.type.2}).

The key point here is that the critical solution is attractive on the
critical manifold. Stated differently, nearly-critical solutions will
experience ``funneling'' effects as all but one mode converge towards
$Z^\star$. If $P \approx P^\star$, then the unstable mode, \ie the
mode ``perpendicular'' to $\mathcal{C}$, will be small until later in
the evolution, thus allowing for the observation of nearly-critical
solutions. In this case, all but one mode of the solution are ``washed
out'' by the interaction with the critical solution, thus explaining
both the universality of the solution and the mass-scaling relation.

\subsubsection{Type-I critical phenomena.} Type-I critical phenomena
are the ones in which the black hole formation turns on at finite mass
and the critical solution presents a non-selfsimilar stationary or
periodic solution configuration. The scaling quantity here is the
lifetime of the metastable solution
\begin{equation}\label{eq:criphen.type1.fond.relation}
t_p = - \frac{1}{\lambda} \ln | P - P^\star | + \mathrm{const}\, ,
\end{equation}
where $\lambda$ does not depend on the initial data. This scaling can
be justified using simple arguments similar to the ones presented in
\cite{gundlach_2007_cpg} for the mass scaling in the type-II case.
\section{Numerical setup}\label{sec:numerical.setup}

In what follows we briefly describe the numerical setup used in the
simulations and the procedure followed in the construction of the
initial data. In essence, we use the \texttt{Whisky2D} code described
in detail in~\cite{Kellermann:08a} and based on the 3-dimensional code
\texttt{Whisky}~\cite{Baiotti04,Giacomazzo:2007ti,Baiotti08}, to solve
numerically and in 2 spatial dimensions the full set of Einstein
equations
\begin{equation}\label{eq:field.equations}
	G_{\mu\nu} = 8 \pi T_{\mu\nu}\,,
\end{equation}
where $G_{\mu\nu}$ is the Einstein tensor and $T_{\mu\nu}$ is the
stress-energy tensor. More specifically, we evolve a
conformal-traceless ``$3+1$'' formulation of the Einstein equations as
presented in~\cite{pollney:2007ss}, in which the spacetime is
decomposed into 3D spacelike slices, described by a metric
$\gamma_{ij}$, its embedding in the full spacetime, specified by the
extrinsic curvature $K_{ij}$, and the gauge functions $\alpha$ (lapse)
and $\beta^i$ (shift), which specify a coordinate frame.  Axisymmetry
is imposed using the ``cartoon'' technique~\cite{Alcubierre99a} and
the equation are solved using finite differencing of order three. The
chosen slicing condition is the popular ``$1+\log$'' while the chosen
spatial-gauge is the Gamma-freezing one. The field equations for the
three-metric $\gamma_{ij}$ and the second fundamental form $K_{ij}$
are coupled with the equations of motion of general relativistic
hydrodynamics
\begin{equation}\label{eq:hydro.equations}
	\nabla_\mu (\rho u^\mu) = 0\,, \qquad \nabla_\nu T^{\mu\nu} = 0\,,
\end{equation}
where $\rho$ is the (rest) baryonic mass density, $u^\mu$ is the four-velocity
of the fluid and $T^{\mu\nu}$ is the stress-energy tensor of a perfect fluid
\begin{equation}
T_{\phantom{\mu}\nu}^\mu = \rho H u^\mu u_\nu + p \delta_{\phantom{\mu}\nu}^\mu\,.
\end{equation}
Here, $H \equiv 1 + \epsilon + p / \rho$ is the specific enthalpy, $p$
is the pressure, $\delta_{\phantom{\mu}\nu}^\mu$ is the Kronecker
delta and $\epsilon$ is the specific internal energy so that $e = \rho
( 1 + \epsilon)$ is the energy density in the rest-frame of the
fluid. These equations are closed using an ideal-gas equation of state
$p = (\Gamma - 1) \rho \epsilon$, with adiabatic exponent
$\Gamma=2$. The solution of relativistic hydrodynamics equations is
obtained via a conservative formulation of (\ref{eq:hydro.equations})
as discussed in~\cite{Kellermann:08a} and the use of high-resolution
shock-capturing (HRSC) schemes with a piecewise parabolic method (PPM)
for the reconstruction of the primitive variables. The time-stepping
is done with a third-order total-variation diminishing Runge-Kutta
algorithm. Finally, the spatial discretization is done on a uniform
grid having resolution of either $h=0.1$ (medium resolution) or
$h=0.08$ (high resolution). The outer boundary of the computational
domain is set at $R=15$ and we have verified that the proximity of the
outer boundary does not influence significantly the critical solution.

The equilibrium configuration curves in the $(\rho,M_{\mathrm{ADM}})$ plane and
the perturbative oscillations frequencies quoted in the text have been computed
using two codes kindly provided to us by S'i. Yoshida~\cite{Yoshida01} and C.
Chirenti~\cite{chirenti_2007_htg}.

\begin{table}
\caption{\label{table1}Properties of some of the representative models
  considered and shown either in figure~\ref{fig:initial.data} or in
  figures~\ref{fig:sub.critical.final.fate}
  and~\ref{fig:three.solutions}. More specifically, $N_1$ and $S_1$
  are the extremes of the range of central densities considered, $P_1$
  is a largely subcritical model which expands to models $P_2$--$P_4$
  as the resolution is increased, while $Q_1$ and $R_1$ represent the
  closest super and subcritical approximation of the critical
  solution, respectively.}
	\begin{indented}
	\item[]
	\begin{tabular}{lllllcc}
		\br
Point & $\rho_c$ & $K$ & $M_{\mathrm{ADM}}$ & $M_b$ & subcritical & supercritical \\
\mr
$N_1$ & $0.00395000$ & $71.77$ & $1.3879$ & $1.5194$ & $\surd$ & $-$     \\
$P_1$ & $0.00459316$ & $71.39$ & $1.3832$ & $1.5194$ & $\surd$ & $-$     \\
$P_2$ & $0.00341517$ & $72.23$ & $1.3754$ & $1.5077$ & $\surd$ & $-$     \\
$P_3$ & $0.00378525$ & $71.58$ & $1.3788$ & $1.5134$ & $\surd$ & $-$     \\
$P_4$ & $0.00387685$ & $71.61$ & $1.3809$ & $1.5161$ & $\surd$ & $-$     \\
$Q_1$ & $0.00459322$ & $71.39$ & $1.3832$ & $1.5194$ & $-$     & $\surd$ \\
$R_1$ & $0.00459322$ & $71.39$ & $1.3832$ & $1.5194$ & $\surd$ & $-$     \\
$S_1$ & $0.00508840$ & $71.95$ & $1.3842$ & $1.5194$ & $-$     & $\surd$ \\
\br
\end{tabular}
\end{indented}
\end{table}

\subsection{Initial Data}

The initial data consists of a family of spherical stars having fixed
baryonic mass
\begin{equation}
	\label{eq:total.baryonic.mass} M_b = 1.5194\equiv {\bar M}_b\,,
\end{equation}
constructed using a polytropic equation of state $p = K \rho^\Gamma$,
with $\Gamma=2$. Each model is computed by fixing its central
rest-mass density, $\rho_c$, while the value of $K$ is fixed after
imposing the condition (\ref{eq:total.baryonic.mass}). The reason for
this choice is that we want to guarantee that all the models
considered have, at least initially, the same baryonic mass to the
precision in expression~(\ref{eq:total.baryonic.mass}). Solutions with
different baryonic mass, in fact, are effectively in different phase
spaces and thus not useful when looking at a critical behaviour. Of
course different models will also be slightly different because the
perturbations will slightly alter their mass-energy or because
although $M_b$ is conserved to high precision by employing a
conservative formulation of the equations, it is nevertheless not
conserved to machine precision. All of these latter errors, however,
are entirely resolution dependent and can, therefore, be singled out
by considering simulations at different resolutions.

These initial models have been evolved under the sole effects of the
perturbations induced by the truncation error. Besides depending on
resolution (and converging away), the amplitude of these perturbations
is difficult to measure as it depends on a number of different sources
of error, such as the interpolation error of the one-dimensional
initial data on the three-dimensional Cartesian grid, or the treatment
of low density ``atmosphere'' regions, which are not measurable
directly. However, an indirect measure can be obtained by looking at a
short evolution of a stable spherical star which, in absence of any
numerical error, would not exhibit any dynamics but which, in
practice, oscillates under the effects of these
perturbations~\cite{Font02c, Baiotti04, Duez05MHD0, Giacomazzo:2007ti,
  Anderson2007,Kellermann:08a,liebling_2010_emr}. The amplitude of the
observed oscillations can be then interpreted as an indirect measure
of the numerical perturbation. In particular, we can consider the
value of the average velocity in the radial direction during the first
iterations as an estimate of the amplitude of an equivalent velocity
perturbation. In this case, for a spherical star with $\rho_c =
0.00128$ and $K=100$ evolved for $100$ timesteps on a $h=0.1$ grid, we
measure an average velocity, $v^r \simeq 1.1 \times 10^{-5}$. Further
insight can also be gained by the average of the momentum constraint
violation in the radial direction and the Hamiltonian constraint
violation, which we measure to be $\simeq 2.3\times 10^{-7}$ and
$\simeq 6.1 \times 10^{-6}$, respectively.

The determination of the critical value of the central density
$\rho_c^\star$ is obtained rather straightforwardly via a
bisection-like strategy within the initial interval
\begin{equation}
\label{eq:initial.data.density}
0.00395 < \rho_c < 0.0050884\,,
\end{equation}
where the extrema correspond to a stable oscillating star or to one
collapsing promptly to a black hole, respectively.

\begin{figure}
\begin{center}
\includegraphics[width=8.0cm,angle=0]{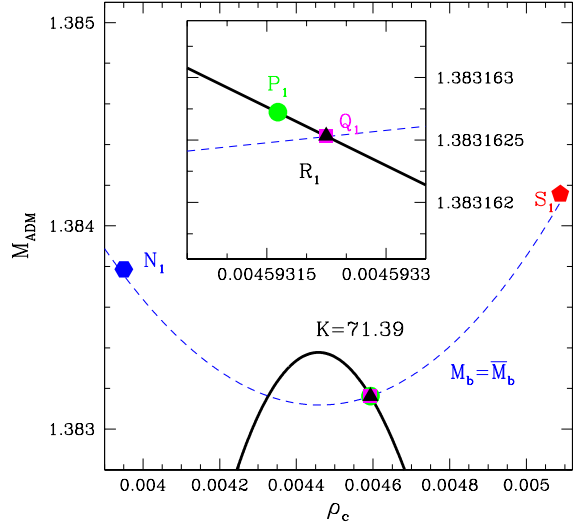}
	\vskip -0.5cm
	\caption{Position of some of the most important models in the
          $(\rho_c,M_{\mathrm{ADM}})$ plane, where the solid (black)
          line refers to a sequence with $K=71.39$, while the dashed
          (blue) line refers to a sequence of models having
          baryonic.mass $M_b = 1.5194 = {\bar M}_b$. The points $N_1$
          and $S_1$ are the extremes of the range of central densities
          considered [\cf eq.~(\ref{eq:initial.data.density})], $P_1$
          is a largely subcritical model, while $Q_1$ and $R_1$
          represent the closest super and subcritical approximation of
          the critical solution, respectively. The inset shows a
          magnification of the region near the critical solution; the
          properties of the model are reported in table~\ref{table1}.
          \label{fig:initial.data}}
	\end{center}
\end{figure}

The main properties of the initial data are collected in
table~\ref{table1} and summarized in figure~\ref{fig:initial.data},
which reports the position of some of the most important models
discussed in this paper in the $(\rho_c,M_{\mathrm{ADM}})$ plane. More
specifically, $N_1$ and $S_1$ are the extremes of the range of central
densities considered [\cf eq.~(\ref{eq:initial.data.density})], $P_1$
is a largely subcritical model which expands to models $P_2$--$P_4$ as
the resolution is increased (\cf
figure~\ref{fig:sub.critical.final.fate}), while $Q_1$ and $R_1$
represent the closest super and subcritical approximation of the
critical solution, respectively. Note that $R_1$ and $Q_1$ differ only
by the $4.6\times 10^{-8}\ \%$ in the central density and thus they
appear identical in the figure. Note also that $P_1$, $Q_1$ and $R_1$
are all on the unstable branch of the models of equilibrium and are
therefore linerarly unstable.

As a final remark we note that although the use of an axisymmetric
system of equations is not strictly necessary for the
spherically-symmetric initial data considered here, their numerical
solutions in 2 spatial dimensions via the \texttt{Whisky2D} code has
been useful in view of the connections between the critical behaviour
discussed here and the one presented in the companion
paper~\cite{Kellermann:10}, where the head-on collision of equal-mass
neutron stars is considered. The possibility of using the same
numerical infrastructure and comparable truncation errors has been in
fact very important in determining the connections between the two
critical behaviours.

\section{Results}\label{sec:results}

In what follows we discuss the nonlinear dynamics of the spherical
stars as these evolve away from their initial state on the unstable
branch and exhibit a critical behaviour.

\subsection{Critical solution}
\label{critical_solution}

\begin{figure}
\includegraphics[width=6.5cm]{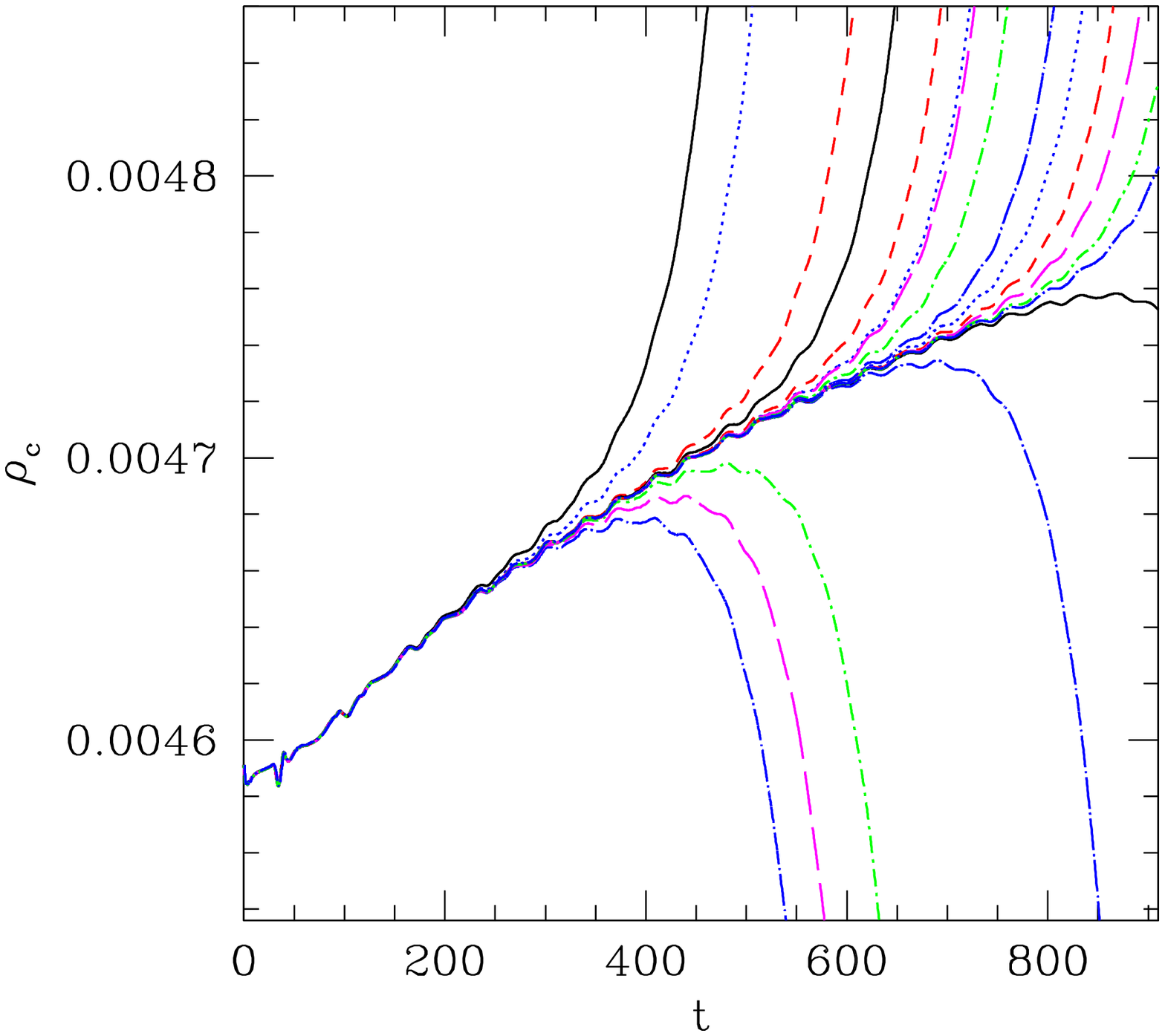}
\hskip 0.5cm
\includegraphics[width=6.5cm]{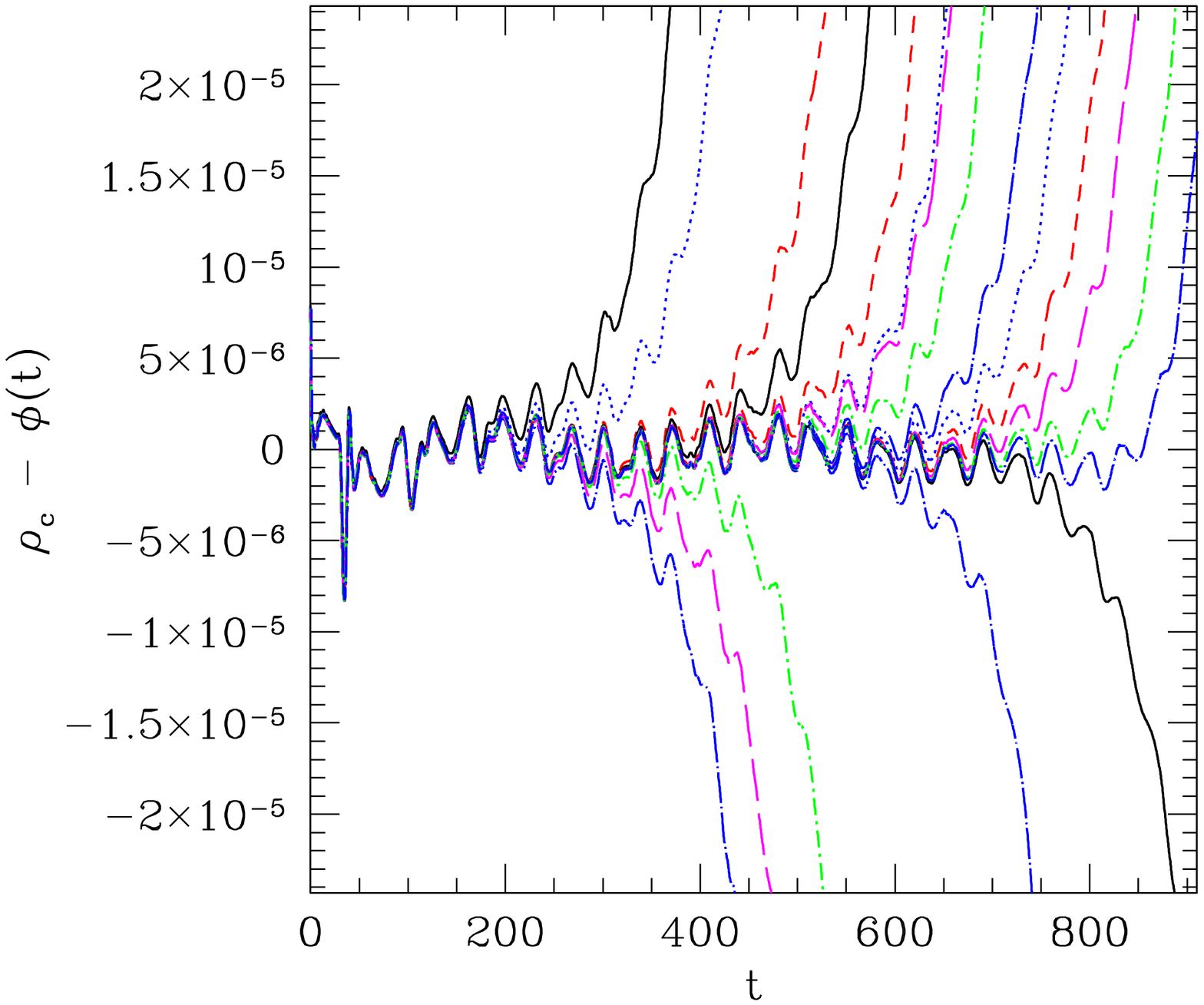}
\vskip -1.0cm
	\caption{Left panel: evolution of the central rest-mass
          density near the critical threshold with different lines
          referring to different initial models.  Right panel: the
          same as in the left panel but corrected for the secular
          evolution given by
          eq.~(\ref{eq:cent.dens.fit}).\label{fig:cent.dens.stdres}}
\end{figure}

We first consider the evolution of models in the window
(\ref{eq:initial.data.density}) under the sole effect of the
numerically-induced perturbations. Some of these models, namely the
supercritical ones, collapse to black hole, while others, namely the
subcritical ones, undergo a sudden expansion followed by a relaxation
towards the corresponding model on the stable branch of the spherical
star solutions. This is clearly shown in
figure~\ref{fig:cent.dens.stdres}, which reports the evolution of the
central rest-mass density and where different lines refer to different
initial data in the interval
\begin{equation}
\label{rho_range}
0.0045931640625  \leq \rho_c \leq 0.00459371875 \,.
\end{equation}

By looking at left panel figure~\ref{fig:cent.dens.stdres} it is quite
apparent how the survival time of the metastable solution increases as
the initial models approach the critical threshold and both the
subcritical and the supercritical solutions overlap for a long part of
the evolution, before departing exponentially. It is also worth
remarking that the linear stability analyses of theses models
indicates that they are linearly unstable with a characteristic
collapse time (\ie the inverse of the imaginary part of the complex
eigenfrequency of the fundamental mode) $\tau \simeq 440$.  Yet, as
shown in figure ~\ref{fig:cent.dens.stdres}, the metastable models
survive for much longer times and for almost $\tau \simeq 850$ for the
models closest to the critical threshold.

A similar behaviour in the evolution of the central rest-mass density
has been observed also in the simulations reported
in~\cite{liebling_2010_emr}, although those refer to magnetized and
rotating stellar models and thus, being them result of
three-dimensional simulations, are restricted to a much smaller
interval of significant figures. In addition, and as mentioned in the
Introduction, evidence for a type-I critical behaviour for the
evolution of the central rest-mass density has been shown also in the
head-on collision of two equal-mass spherical stars~\cite{Jin:07a} and
will be further discussed in the companion paper~\cite{Kellermann:10}.

\begin{figure}
\begin{center}
	\includegraphics[width=8.0cm,angle=0]{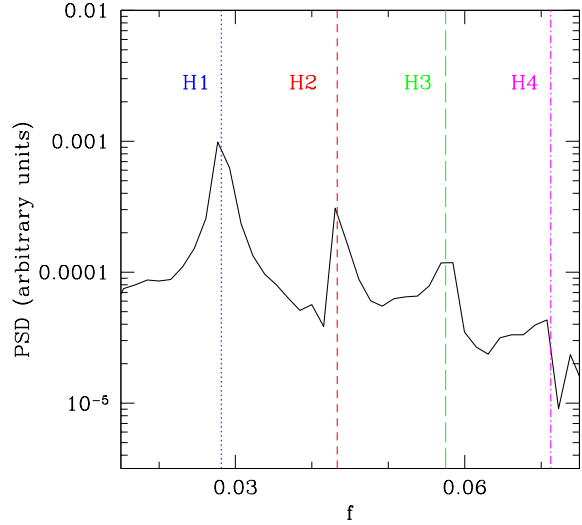}
	\vskip -0.8cm
	\caption{Power spectral density of the evolution of the
          central rest-mass density for the model closest to the
          putative critical solution (\ie with
          $\rho_c=0.0045932248034$) when the secular drift part
          (\ref{eq:cent.dens.fit}) has been removed from the data. The
          eigenfrequencies associated with the corresponding spherical
          star model are also shown as vertical
          lines.\label{fig:critical_solution.spectrum}}
\end{center}
\end{figure}

As the secular evolution in the central density is a well-known
``feature'' of the numerical solution of relativistic multidimensional
stellar models and has been observed in codes implementing very
different numerical methods and formulations of the Einstein
equations~\cite{Font02c, Baiotti04, Duez05MHD0, Giacomazzo:2007ti,
  Anderson2007,Kellermann:08a,liebling_2010_emr}, we have isolated
this secular behaviour by computing a least-square fit of the common
part of the evolution in order to isolate the true dynamics from the
low-frequencies numerical components. More specifically, we have
modeled the evolution of the central rest-mass density of the
metastable equilibrium via the \textit{Ansatz}
\begin{equation}\label{eq:cent.dens.fit}
\phi(t) = \rho_0 + \rho_1 t + \rho_2 \cos(2\pi h_1 t + \varphi_1) +
	\rho_3 \cos(2\pi h_2 t + \varphi_2)\,,
\end{equation}
where $\rho_0-\rho_2$ are just coefficients in the interpolation and
do not have a particular physical meaning. On the other hand, the
frequencies $h_1$ and $h_2$ are chosen as the two smallest frequencies
appearing in the Fourier spectrum of the central density during the
metastable phase (\cf figure~\ref{fig:critical_solution.spectrum} and
see also discussion below on the spectral power density of the
putative critical solution). The residuals after the fit are shown in
the right panel of figure~\ref{fig:cent.dens.stdres} and help
considerably in appreciating the dynamics of the unstable models near
the critical value.

Using a large set of simulations with resolution of $h=0.1$ and a
straightforward bisection strategy we have located the critical
threshold to black-hole formation at a central density
\begin{equation}\label{eq:critical.central.density}
\rho_c^\star = 0.004593224802 \pm 2.1 \times 10^{-12} \,.
\end{equation}
Clearly, we expect this value to depend on the initial perturbation
and thus on the resolution used, as well as on the numerical method
employed. On the other hand, we also expect that the associated
solution and the critical exponent to be ``universal'', in the sense
that they should not depend depend sensitively on the perturbation or
on the particular family of initial data as far as this family is
characterized by a single parameter and thus intersects the critical
manifold $\mathcal{C}$ in a single point which is near enough to this
solution. In this case, in fact, the associated critical solution is
supposed to be at least locally attractive on a sub-manifold of the
phase space of codimension one.

\begin{figure}
\includegraphics[width=6.25cm]{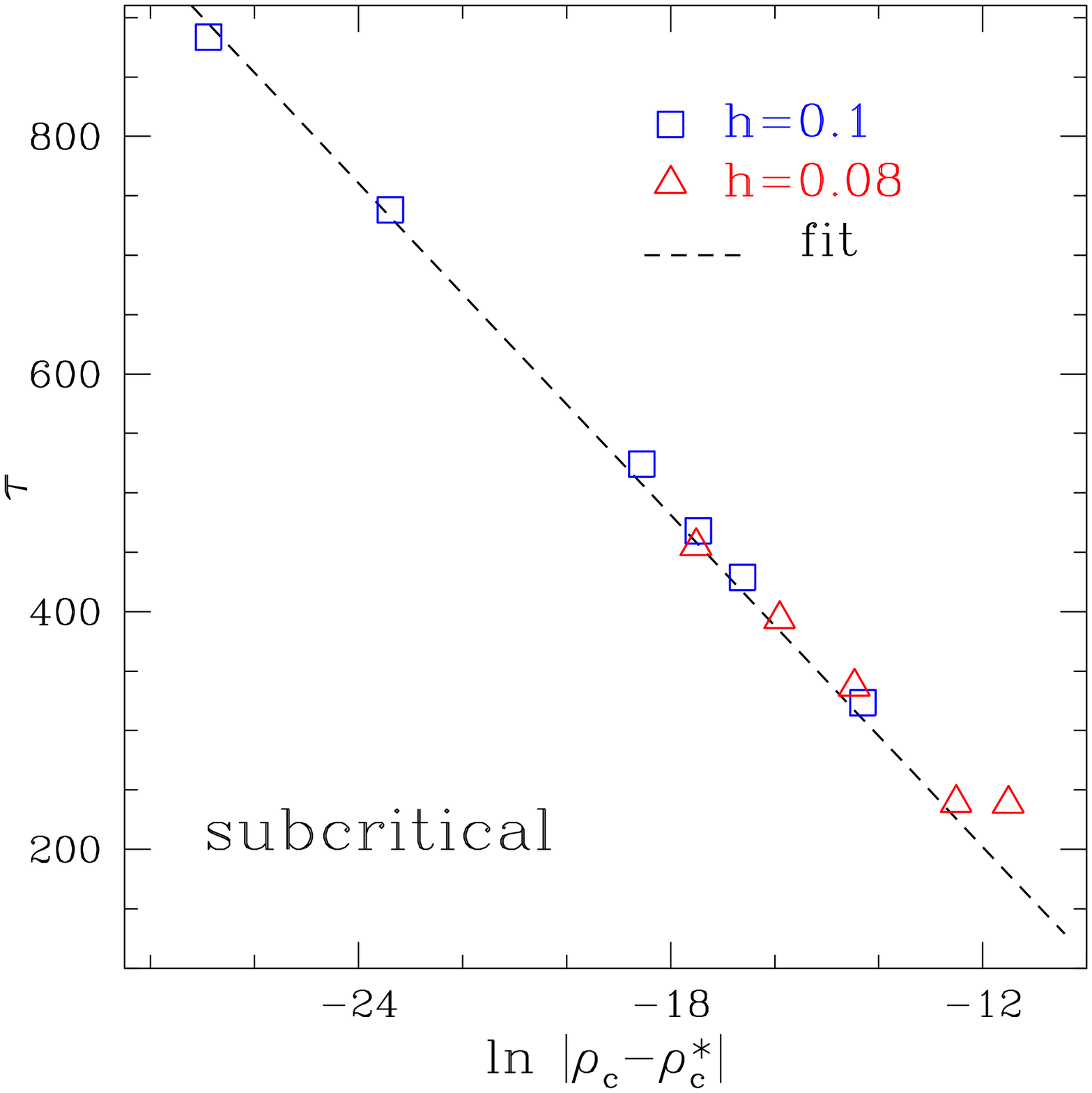}
	\hskip 0.5cm
\includegraphics[width=6.25cm]{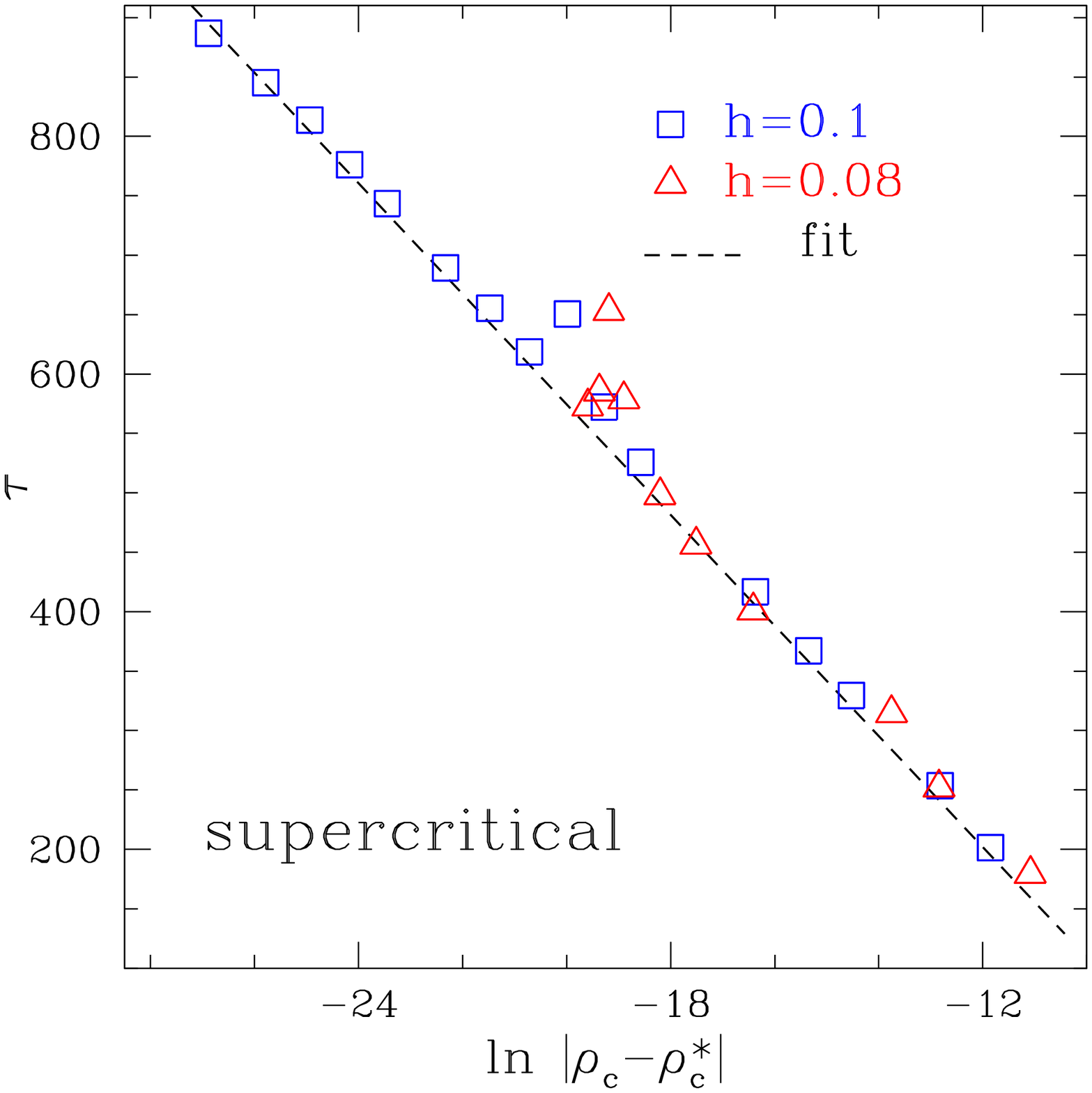}
	\caption{Escape time $\tau$ as a function of $\ln | \rho -
          \rho^\star|$ for subcritical (left panel) and supercritical
          solutions (right panel), respectively. The blue squares
          corresponds to the results obtained with the $h=0.1$
          resolution, while the red triangles to the results obtained
          with the $h=0.08$ resolution. The dashed lines represent the
          fit obtained using (\ref{eq:tp.scaling}) with the $\lambda$
          obtained from the $h=0.1$ solutions.\label{fig:escape.time}}
\end{figure}

To validate that the behaviour discussed so far and shown in
figure~\ref{fig:cent.dens.stdres} does represent a type-I critical
behaviour we compute the survival time of the metastable solution
$\tau$, \ie the \textit{``escape time''}, and study how this varies as
the critical solution is approached. We recall that we expect that the
escape time near the critical for a type-I critical phenomena should
behave as
\begin{equation}\label{eq:tp.scaling}
\tau = - \frac{1}{\lambda} \ln | \rho_c - \rho_c^\star | + \mathrm{const}\,,
\end{equation}
and such expected solution is indeed shown as a dashed line in
figure~\ref{fig:escape.time}. Also shown with squares and triangles
are the computed escape times for different initial data and different
resolutions (blue squares for $h=0.1$ and red triangles for $h=0.08$).
The latter are calculated in terms of the time $\tau_\epsilon$ at
which the relative difference between the observed central baryonic
density and the best approximation of the critical
solution $\phi(t)$~(\ref{eq:cent.dens.fit}) becomes larger than
$\epsilon$. We find that, for a large enough $\epsilon$, such that $1
\gg \epsilon > \epsilon^\star > 0$, these times depend only weakly on
$\epsilon$ and thus give a good measure of the departure time from the
critical solution. A value of $\epsilon = 0.5\%$ provides a
sufficiently accurate measure and this is the one employed for the
data points shown in figure~\ref{fig:escape.time}. We finally estimate
the critical exponent $\lambda$ by making a linear least-square
regression of the data points of sub- and supercritical solutions and
then by taking the average of the two values. Using the
medium-resolution  $h=0.1$ simulations we therefore obtain for the
critical exponent
\begin{equation}
\label{crit_exponent}
\lambda = 0.02149665\,,
\end{equation}
with a coefficient of determination $R^2$ relative to the linear
regression (\ref{eq:tp.scaling}) and computed on the full dataset
containing both sub and supercritical solutions, of $0.960517$. The
critical exponent~(\ref{crit_exponent}) is found also in the case of
the $h=0.08$ simulations, although in this case the scattering is
somewhat larger and the data agrees within $7\%$. We note that these
high-resolution simulations are computationally very expensive and
this is why we have restricted them to a smaller set of initial
data. Clearly, the match between the computed escape times and the one
expected from the critical behaviour is very good over the 6 orders of
magnitude in $|\rho_c - \rho_c^\star|$ spanned by our data-set and
thus provide convincing evidence that indeed critical behaviour can be
found in the dynamics of linearly unstable spherical stars.

As a final remark we note that while the evidence for a critical
behaviour is clear, much less clear is the physics of the critical
solution which is, after all, a perturbed spherical star. Recent
studies of nonlinear perturbations of relativistic spherical stars
have shown that linearly unstable stars can be stabilized via
nonlinear couplings among higher-order modes~\cite{Gabler:2009yt}. It
is possible that such a nonlinear coupling is present also here and we
conjecture therefore that the stability of the metastable solution is
due to mode coupling of the first overtones of the fundamental
mode. Support to this conjecture comes from the power spectral density
in figure~\ref{fig:critical_solution.spectrum}, which shows that,
apart from the $F$-mode which is obviously missing as it has only
imaginary eigenfrequency, the spectrum of the metastable solution is
essentially identical to the one of an excited spherical star with
$(\rho_c,K) = (\rho^\star,K^\star)$ and $M_b={\bar
  M}_b$. Interestingly, most of the energy is in the first overtone,
$H1$, even though the numerical perturbation can be thought as ``white
noise'' exciting all the modes of the star with almost equal
energy. The behaviour discussed above persists also when considering
models with higher spatial resolutions.

\subsection{Subcritical solutions}

While the final fate of supercritical solutions is clearly that of
leading to a collapse and to the formation of a black hole, the one of
subcritical solutions deserves a more detailed explanation. As one
would expect, given that the initial data represent linearly unstable
stars, the subcritical solutions show a first expansion as the star
migrates to the stable branch of the equilibrium configurations, which
is then followed by a slow relaxation where the central rest-mass
density exhibits strong oscillations around smaller and smaller
values, that would eventually reach in the continuum limit, the value
corresponding to the model on the stable branch having the same
gravitational mass of the initial one. In practice, however, the
migration to the stable branch is accompanied small losses both in the
gravitational mass and in the rest-mass which, although smaller than
$\simeq 0.7\%$, need to be taken properly into account.

More specifically, we have analyzed in detail the evolution of the
largely subcritical model $P_1$, (\cf table \ref{table1}), which is an
unstable spherical star with an F-mode whose imaginary part of the
eigenfrequency is $\nu_i= 0.461\, \mathrm{kHz}$.  We evolve therefore
evolved such a model it with three different spatial resolutions of $h
= 0.1$, $h=0.09$ and $h=0.08$, and studied its migration to the stable
branch. The asymptotic state of the solution and in particular to the
final central rest-mass density $\rho_f$ is estimated by modelling the
time evolution of the oscillating star on the stable branch with a
simple \textit{Ansatz} of the type $\rho(t) = \rho_f + {\rho_1}/{t} $
and by performing a nonlinear least square fit on an appropriate
window including the final part of the dynamics.  For any given
resolution we have then computed the total baryonic-mass losses due to
the numerical dissipation $\Delta M_b = M_b - M_{b,f}$, and determined
the polytropic coefficient $K_f$ yielding a spherical stellar model
with central rest-mass density $\rho_f$ and baryon mass
$M_{b,f}$. Clearly, for such a model it is then also possible to
compute the gravitational mass and thus track the migration on a
$(\rho_c\,, M_{_{\rm ADM}})$ plane.

\begin{figure}[h]
\begin{center}
\includegraphics[width=8.0cm,angle=0]{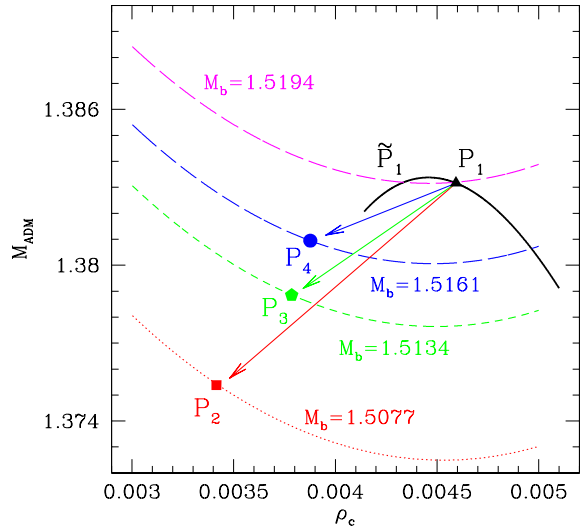}
\vskip -0.5cm
\caption{Dynamics of the migration on a $(\rho_c\,,M_{_{\rm ADM}})$
  plane. The linearly unstable and largely subcritical model $P_1$
  migrates to a new solution on the stable branch of equilibrium
  configurations. Indicated with $P_2-P_4$ are the new asymptotic
  states for resolutions $h=0.1-0.08$, respectively. Indicated with a
  thick solid line is the sequence of initial models having the same
  polytropic index of $P_1$, while indicated with dotted and dashed
  lines are sequences of models having the same rest-mass as the
  asymptotic models $P_2-P_4$. Finally, shown as $\tilde{P}_1$ is the
  asymptotic state of $P_1$ in the continuum limit; note that even for
  the coarse-resolution case the changes in baryonic and rest-mass are
  only of $\simeq 0.7\%$.\label{fig:sub.critical.final.fate}}
	\end{center}
\end{figure}

The overall results of these migrations are shown in
figure~\ref{fig:sub.critical.final.fate}, where we report the stellar
configurations on $M_b = \mathrm{const.}$ curves. The minimum of each
curve corresponds to the maximum in the usual
$(\rho,M_{\mathrm{ADM}})$, $K= \mathrm{const.}$, plots and separates
the stable and unstable branches of solutions. When a resolution of
$h=0.1$ is used the model $P_1$ migrates to the new asymptotic model
$P_2$, while it will migrate to models $P_3$ and $P_4$ as higher
resolutions of $h=0.09$ and $h=0.08$ are used, respectively. Note that
already with the coarsest resolution of $h=0.1$ the losses in
gravitational masses are $\simeq 0.65\%$ and that these decrease to
$\simeq 0.16\%$ when a resolution of $h=0.08$ is used.  Finally,
indicated with $\tilde{P}_1$ is the expected asymptotic model when the
numerical losses are extrapolated to the continuum limit\footnote{Note
  that we do not mark this point with a symbol as it does not
  correspond to a numerically computed value, as it instead for $P_2$,
  $P_3$ and $P_4$}; clearly, in the limit $h\to 0$, the migration of
model $P_1$ takes place to a new state having the same gravitational
and baryonic mass as the initial one.

\subsection{Perturbation of nearly-critical solutions}

\begin{figure}[ht]
\begin{center}
\vskip 0.5cm
	\includegraphics[scale=0.375]{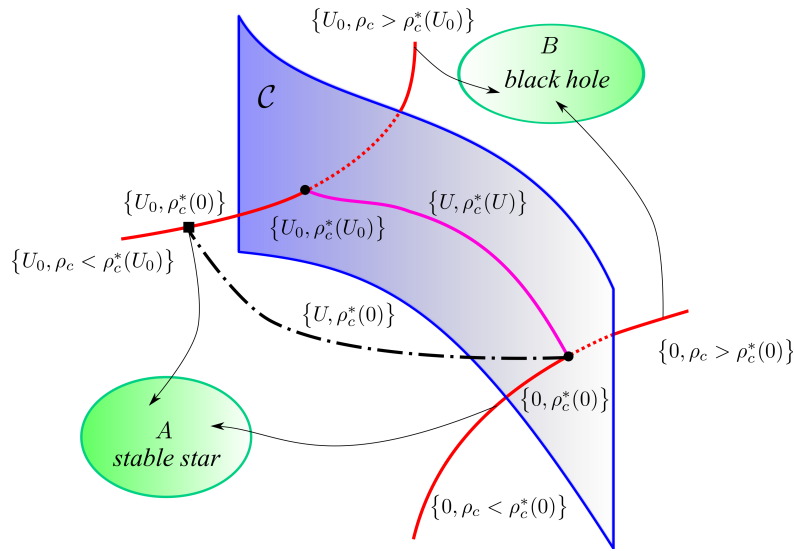}
\end{center}
	\caption{\label{fig:phase.space.perturb}Phase-space diagram
          representation of nearly-critical solutions. In particular,
          we show with red solid curves two one-parameter families of
          initial data, perturbed by the addition of a radial velocity
          profile in the form (\ref{eq:velocity.perturbation}) with $U
          = 0$ or $U = U_0 > 0$. The locus of the critical points,
          $\{U,\rho_c^\ast(U)\}$, is shown with a violet solid line,
          while the family of initial-data $\{U,\rho_c^\ast(0)\}$ is
          shown with a black dot-dashed line and the point
          $\{U_0,\rho_c^\ast(0)\}$ is marked with a filled square. The
          latter represents therefore the family of initial data
          obtained by adding a velocity perturbation with amplitude
          $U$ to the model with central density would when $U=0$. Also
          highlighted with filled circles are the critical points for
          the families with $U=0$ and $U=U_0$, \ie $\{0,
          \rho_c^\ast(0)\}$ and $\{U_0, \rho_c^\ast(U_0)\}$.}
\end{figure}

As discussed in Sect.~\ref{critical_solution}, the central rest-mass
density of the linearly unstable models can be used as a critical
parameter for the gravitational collapse of a linearly unstable
spherical star, in contrast to what has been observed for example by
Novak in~\cite{novak_2001_vic} or by Noble in~\cite{Noble08a}. We
believe this is due to the very different set of initial data selected
here and in~\cite{novak_2001_vic,Noble08a}. Indeed, the reason why
this behaviour has not been observed in many previous studies is that
we consider initial stellar models that are already linearly
\textit{unstable}, in contrast with what done
in~\cite{novak_2001_vic,Noble08a}, where the initial models are
instead linearly \textit{stable} and then subject to a perturbation
(either by introducing a radial
velocity~\cite{novak_2001_vic,Noble08a}, or by considering employing
the interaction with a scalar field~\cite{noble_2003_nsr}). For our
set of initial data, therefore, the critical solution is essentially a
spherical star with an unstable F-mode, and any finite perturbation
exciting this mode will change the solution in a dramatic way (A
discussion of this change within a phase-space description will be
made later on when presenting figure
\ref{fig:phase.space.perturb})\footnote{With ``perturbation'' we are
  here referring to a globally coherent, resolution independent
  perturbation such as the one given in
  eq.~(\ref{eq:velocity.perturbation}). This has to be contrasted with
  the random, truncation-error induced and resolution-dependent
  perturbations we have considered in
  Sect.~\ref{critical_solution}}.

To confirm this hypothesis, we follow~\cite{novak_2001_vic} and
\cite{Noble08a}, and construct a new family of spherical initial data
obtained by perturbing the slightly supercritical model $Q_1$ (\cf
table \ref{table1}) via the addition of a radial velocity perturbation
in the form of the 3-velocity component
\begin{equation}\label{eq:velocity.perturbation}
v^r(x) = \frac{U}{2} ( 3 x - x^3 ), \qquad x \equiv \frac{r}{R_\star}\,,
\end{equation}
where $U$ is the amplitude of the perturbation at the surface of the
star, $R_\star$ and can be either positive (outgoing radial velocity)
or negative (ingoing radial velocity). Because the
perturbation~(\ref{eq:velocity.perturbation}) matches the
eigenfunction of an idealized F-mode perturbation, it should excite
the only unstable mode of the critical solution.

Performing simulations for different values of $U$ and a resolution
$h=0.1$ we find, not surprisingly, that for negative values of $U$ the
perturbed models of $Q_1$ collapse to a black hole. Furthermore,
because in this case the radial velocity accelerates the development
of the unstable mode, the larger the values of $U$ the shorter the
time to collapse, \ie \mbox{$\tau \sim -c_1 \log(U) + c_2$}, where
$c_1$ and $c_2$ are positive constant coefficients. On the other hand,
for positive values of $U$, the perturbed models of $Q_1$, which we
recall is supercritical for $U=0$, becomes subcritical and shows the
same qualitative behaviour as that of model $R_1$. Hence, a suitably
perturbed supercritical model can behave as a subcritical one.

\begin{figure}[ht]
\begin{center}
	\includegraphics[width=8.0cm]{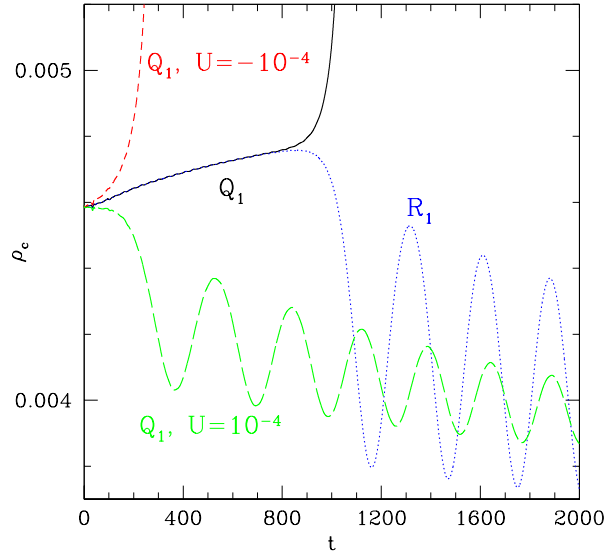}
\end{center}
	\vskip -0.7cm
	\caption{\label{fig:three.solutions} Perturbation of
          nearly-critical solutions. The solid (black) line represents
          the supercritical solution $Q_1$, while the dotted (blue)
          line represents the subcritical solution $R_1$. The dashed
          lines show again the evolution of $Q_1$, but when subject to
          a negative (red short-dashed line) or positive (green
          long-dashed line) velocity perturbation. Clearly, in the
          latter case the supercritical solution $Q_1$ becomes
          subcritical and shows the same behaviour as the solution
          $R_1$.}
\end{figure}

The dynamics of these perturbed, nearly-critical models is shown in
figure~\ref{fig:three.solutions}, where the solid (black) line
represents the supercritical solution $Q_1$, while the dotted (blue)
line represents the subcritical solution $R_1$. The dashed lines show
again the evolution of $Q_1$, but when subject to a positive (red
short-dashed line) or negative (green long-dashed line) velocity
perturbation. The dynamics shown in figure~\ref{fig:three.solutions}
underlines an important characteristic of critical phenomena: the
precise value of the critical parameter at the intersection between
the one-parameter family of solutions and the critical manifold
depends on the family itself. In particular this means that if we fix
a value of the perturbation amplitude, $U \neq 0$, we have to expect
to find the critical solution at a value of $\rho_c^\star(U)$
different from the one quoted in (\ref{eq:critical.central.density})
which is attained in the case $U=0$. For this reason the application
of a non infinitesimal perturbation to a nearly-critical solution
results in a dramatic change in the dynamics of the system.

The phase-space representation of this concept is summarized in
figure~\ref{fig:phase.space.perturb}, where we show two one-parameter
families of perturbed TOV initial data, whose critical parameter,
$\rho_c$, is the central rest-mass density. The perturbation is given
by the composition of truncation errors and of a radial velocity
perturbation $U$ in the form (\ref{eq:velocity.perturbation}), where
$U = 0$ or $U = U_0 > 0$. As these families represent different
initial configurations, they will intersect the critical manifold
$\mathcal{C}$ at two different points, with correspondingly different
values of the critical parameter $\{0,\rho^\ast_c(0)\}$ and
$\{U_0,\rho^\ast_c(U_0)\}$ (these points are marked as filled
circles)\footnote{In our notation, the point $\{U_0,
  \rho^\ast_c(U_0)\}$ is the critical solution with initial velocity
  perturbation given by (\ref{eq:velocity.perturbation}) with
  $U=U_0$. Similarly, a configuration $\{U_0, \rho^\ast_c(0)\}$ will
  be a member of the family with initial velocity perturbation $U_0$,
  but with a central density which is the critical one for a model
  with $U=0$}. In particular, when $U$ runs between $0$ and $U_0$, the
set of critical configurations $\{U,\rho^\ast_c(U)\}$ will represent a
curve on the critical manifold $\mathcal{C}$ and this is shown with a
violet solid line in figure~\ref{fig:phase.space.perturb}. Considering
now a configuration near $\{0,\rho^\ast_c(0)\}$ and applying to it a
velocity perturbation in the form (\ref{eq:velocity.perturbation})
with $U = U_0$, will produce a new configuration
$\{U_0,\rho^\ast_c(0)\}$ which is not necessarily on the critical
manifold (this is marked with a filled square). Indeed, the whole
family $\{U,\rho^\ast_c(0)\}$, that is the set of configurations with
a nonzero initial velocity perturbation but central density which is
the critical one for the zero-velocity case, are in general expected
to be outside the critical domain. The family $\{U,\rho^\ast_c(0)\}$
is shown with a black dot-dashed line in
figure~\ref{fig:phase.space.perturb}.

As a final remark we note that another important difference between
the work presented here and that in~\cite{novak_2001_vic,Noble08a} is
that the we find evidence of a type-I critical behaviour with a
periodic solution, in contrast to what found
in~\cite{novak_2001_vic,Noble08a}, which is instead of type-II and
with DSS solutions. We believe the origin of this important difference
and of the presence of a periodic solution is in our use of an
ideal-fluid EOS and hence in the presence of an overall scale in the
problem.  Conversely, the spherical stars considered in the above
mentioned works were evolved using either an ultrarelativistic
EOS~\cite{Hara96a} (which, as commented in the Introduction, are
intrinsically scale-free) or with very strong
perturbations~\cite{novak_2001_vic,Noble08a}, thus in a regime of the
EOS which is approximatively
ultrarelativistic~\cite{gundlach_2007_cpg}.

\section{Conclusions}\label{sec:conclusions}

In general, critical phenomena in gravitational collapse are of great
interest because they play a central role in phase transition of
families of solutions in general relativity. In more specific context
of the dynamics of NSs, type-I critical phenomena have seen a renewed
interested when it was shown that a critical behaviour of this type is
produced in the in head-on collision of NSs~\cite{Jin:07a} or in the
dynamics of rotating magnetized stars~\cite{liebling_2010_emr}.  With
the goal of studying in more detail the occurrence of type-I critical
collapse in NSs, we have therefore employed the 2D general
relativistic code \texttt{Whisky2D} to study a large set of spherical
stellar models having a constant baryon mass. Differently from what
done before by other authors, \eg~\cite{novak_2001_vic,Noble08a}, we
have considered stellar models that are on the ``right'' branch of the
models of equilibrium and thus linearly \emph{unstable}.

Using a simple ideal-fluid EOS and very small perturbations which are
entirely induced by the truncation error, we have found that our
family of initial data exhibits a clear type-I critical behaviour at
at a threshold central rests-mass density of $\rho_c^\star =
0.004593224802 \pm 2.1 \times 10^{-12}$ and with a critical exponent
$\lambda = 0.02149665$. These results thus confirm the conclusions
reached by Liebling et al.~\cite{liebling_2010_emr} but also provide a
more quantitative determination of the threshold and of the nature of
the critical scaling. Exploiting in fact the relative simplicity of
our system, we were able carry out a more in-depth study providing
solid evidences of the criticality of this phenomenon and also to give
a simple interpretation of the putative critical solution as a
spherical solution with the unstable mode being the fundamental
F-mode. As a result, we have shown that for any choice of the
polytropic constant, the critical solution distinguishes the set of
subcritical models migrating to the stable branch of the models of
equilibrium from the set of supercritical models collapsing to a black
hole.

Furthermore, we have studied how the dynamics changes when the
numerically perturbation is replaced by a finite-size,
resolution-independent velocity perturbation and show that in such
cases a nearly-critical solution can be changed into either a sub or
supercritical. Finally, the work presented here here is of direct help
in understanding why the critical behaviour shown in the head-on
collision of two neutron stars is indeed of type-I and why it can be
explained simply in terms of the creation of a metastable stellar
model on the unstable branch of equilibrium
solutions~\cite{Kellermann:10}.

\ack It is a pleasure to thank Bruno Giacomazzo and Filippo Galeazzi
for their help and assistance with the \texttt{Whisky2D} code.  We are
also grateful to Giulio Magli and Carsten Gundlach for useful
discussions and Shin'ichiro Yoshida and Cecilia Chirenti for providing
the codes used to compute the spherical static models and their linear
radial oscillation frequencies. This work was supported in part by the
IMPRS on ``Gravitational-Wave Astronomy'', by the DFG grant
SFB/Transregio~7, and by ``CompStar'', a Research Networking Programme
of the European Science Foundation. The computations were performed on
the Damiana cluster at the AEI.

\section*{References}
\bibliographystyle{iopart-num}

\begin{thebibliography}{10}
\expandafter\ifx\csname url\endcsname\relax
  \def\url#1{{\tt #1}}\fi
\expandafter\ifx\csname urlprefix\endcsname\relax\def\urlprefix{URL }\fi
\providecommand{\eprint}[2][]{\url{#2}}

\bibitem{liebling_2010_emr}
Liebling S~L, Lehner L, Neilsen D and Palenzuela C 2010 (\textit{Preprint}
  \eprint{1001.0575v1})

\bibitem{Kellermann:10}
{Kellerman} T, {Rezzolla} L and {Radice} D 2010 {\em Classical and Quantum
  Gravity\/}  {\em submitted}

\bibitem{Choptuik93}
Choptuik M~W 1993 {\em Physical Reviews Letters\/} {\bf 70} 9

\bibitem{evans_1994_cps}
Evans C~R and Coleman J~S 1994 {\em Physical Reviews Letters\/} {\bf 72}
1782--1785

\bibitem{Hara96a}
{Hara} T, {Koike} T and {Adachi} S 1996 (\textit{Preprint}
\eprint{arXiv:gr-qc/9607010})

\bibitem{neilsen_2000_cpp}
Neilsen D~W and Choptuik M~W 2000 {\em Classical and Quantum Gravity\/} {\bf 17}
761
\bibitem{novak_2001_vic}
Novak J 2001 {\em Astronomy \& Astrophysics\/} {\bf 376} 606 (\textit{Preprint}
  \eprint{gr-qc/0107045v1})

\bibitem{noble_2003_nsr}
Noble S~C 2003 {\em A Numerical Study of Relativistic Fluid Collapse\/} Ph.D.
  thesis University of Texas at Austin (\textit{Preprint}
  \eprint{gr-qc/0310116v1})

\bibitem{Noble08a}
{Noble} S~C and {Choptuik} M~W 2008 {\em Physical Review D\/} {\bf 78} 064059
  (\textit{Preprint} \eprint{0709.3527})

\bibitem{Jin:07a}
{Jin} K~J, {Suen} W~M and {et al} 2007 {\em Physical Review Letters\/} {\bf 98}
131101

\bibitem{musco05}
Musco I, Miller J~C and Rezzolla L 2005 {\em Classical Quantum Gravity\/} {\bf
22} 1405--1424

\bibitem{musco09}
{Musco} I, {Miller} J~C and {Polnarev} A~G 2009 {\em Classical and Quantum
  Gravity\/} {\bf 26} 235001 (\textit{Preprint} \eprint{0811.1452})

\bibitem{gundlach_2007_cpg}
Gundlach C and Mart{\'{i}}n-Garc{\'{i}}a J 2007 {\em Living Reviews in
  Relativity\/} {\bf 10} 5

\bibitem{brady_2002_bht}
Brady P, Choptuik M, Gundlach C and Neilsen D 2002 {\em Classical Quantum
Gravity\/} {\bf 19} 6359--6375

\bibitem{Cahill70}
Cahill M~E and McVittie G 1970 {\em Journal Mathematical Physics\/} {\bf II}
1382

\bibitem{wan_2008_das}
Wan M~B, Jin K~J and Suen W~M 2008 (\textit{Preprint} \eprint{0807.1710v2})

\bibitem{gundlach_1999_cpg}
Gundlach C 1999 {\em Living Reviews in Relativity\/} {\bf 2} 4

\bibitem{Gundlach97f}
Gundlach C 1997 {\em Phys. Rev. D\/} {\bf 55} 695 (\textit{Preprint}
  \eprint{gr-qc/9604019})

\bibitem{Kellermann:08a}
{Kellerman} T, {Baiotti} L, {Giacomazzo} B and {Rezzolla} L 2008 {\em Classical
  and Quantum Gravity\/} {\bf 25} 225007 (\textit{Preprint} \eprint{0811.0938})

\bibitem{Baiotti04}
Baiotti L, Hawke I, Montero P~J, L{\"o}ffler F, Rezzolla L, Stergioulas N, Font
  J~A and Seidel E 2005 {\em Phys. Rev. D\/} {\bf 71} 024035 (\textit{Preprint}
  \eprint{gr-qc/0403029})

\bibitem{Giacomazzo:2007ti}
Giacomazzo B and Rezzolla L 2007 {\em Classical Quantum Gravity\/} {\bf 24} S235
  (\textit{Preprint} \eprint{gr-qc/0701109})

\bibitem{Baiotti08}
{Baiotti} L, {Giacomazzo} B and {Rezzolla} L 2008 {\em Physical Review D\/} {\bf
78} 084033 (\textit{Preprint} \eprint{ArXiv e-prints 0804.0594})

\bibitem{pollney:2007ss}
Pollney D, Reisswig C, Rezzolla L, Szil{\'a}gyi B, Ansorg M, Deris B, Diener P,
Dorband E~N, Koppitz M, Nagar A and Schnetter E 2007 {\em Physical Reviews D\/}
{\bf 76} 124002 (\textit{Preprint} \eprint{arXiv:0707.2559})

\bibitem{Alcubierre99a}
Alcubierre M, Brandt S~R, Br{\"u}gmann B, Holz D, Seidel E, Takahashi R and
Thornburg J 2001 {\em International Journal Modern Physics D\/} {\bf 10}
273--289 (\textit{Preprint} \eprint{gr-qc/9908012})

\bibitem{Yoshida01}
Yoshida S and Eriguchi Y 2001 {\em MNRAS\/} {\bf 322} 389

\bibitem{chirenti_2007_htg}
{Chirenti} C~B~M~H and {Rezzolla} L 2007 {\em Classical and Quantum Gravity\/}
  {\bf 24} 4191--4206 (\textit{Preprint} \eprint{0706.1513})

\bibitem{Font02c}
Font J~A, Goodale T, Iyer S, Miller M, Rezzolla L, Seidel E, Stergioulas N,
  Suen W~M and Tobias M 2002 {\em Physical Reviews D\/} {\bf 65} 084024
  (\textit{Preprint} \eprint{gr-qc/0110047})

\bibitem{Duez05MHD0}
Duez M~D, Liu Y~T, Shapiro S~L and Stephens B~C 2005 {\em Physical Reviews D\/}
{\bf 72} 024028

\bibitem{Anderson2007}
Anderson M, Hirschmann E~W, Lehner L, Liebling S~L, Motl P~M, Neilsen D,
  Palenzuela C and Tohline J~E 2008 {\em Physical Reviews D\/} {\bf 77} 024006
  (\textit{Preprint} \eprint{arXiv:0708.2720})

\bibitem{Gabler:2009yt}
Gabler M, Sperhake U and Andersson N 2009 {\em Physical Reviews D\/} {\bf 80}
064012 (\textit{Preprint} \eprint{0906.3088})

\end{thebibliography}
\providecommand{\newblock}{}

\end{document}